\documentclass[%
 reprint,
 amsmath,amssymb,
 aps,
]{revtex4-2}

\usepackage{graphicx}
\usepackage{dcolumn}
\usepackage{bm}
\usepackage{color}

\begin{document}


\title{Active thermodynamic force driven mitochondrial alignment}

\author{Masashi K. Kajita}
\affiliation{
Department of Applied Chemistry and Biotechnology,
Faculty of Engineering, University of Fukui.
}
\affiliation{
Life Science Innovation Center, University of Fukui.
}

\author{Yoshiyuki Konishi}
\affiliation{
Department of Applied Chemistry and Biotechnology,
Faculty of Engineering, University of Fukui.
}
\affiliation{
Life Science Innovation Center, University of Fukui.
}

\author{Tetsuhiro S. Hatakeyama}
\email{hatakeyama@complex.c.u-tokyo.ac.jp}
\affiliation{%
Department of Basic Science, The University of Tokyo.
}%

\date{\today}

\begin{abstract}
Mitochondria are critical organelles in eukaryotes that produce the energy currency ATP.
In nerve axons, mitochondria are known to align at almost regular intervals to maintain a constant ATP concentration, but little is known about the mechanism.
In this letter, we show theoretically that ATP production and ATP-dependent non-directional movement of mitochondria are sufficient for alignment, even in the absence of an explicit repulsive force between them.
This is similar to thermodynamic forces driven by thermal fluctuations, even generated by non-equilibrium processes, and demonstrates the diversity of mechanisms governing the motion of biological matter.
\end{abstract}

\maketitle



Understanding how the position of organelles is regulated in eukaryotic cells will be important both biologically and physically.
In particular, studying the positioning of mitochondria will be necessary when considering the energetics of the cell \cite{morris_regulation_1993, miller_axonal_2004, sheng_mitochondrial_2012, matsumoto_intermitochondrial_2022}.
The mitochondrion is a fundamental organelle in most eukaryotic cells that produces adenosine triphosphate (ATP) \cite{mitchell_coupling_1961, alberts_molecular_2022}.
ATP is hydrolyzed and used as an energy source for many processes in the cell, such as the synthesis of biomolecules, signal transduction, and the motility of molecular motors, and then it is essential to transport ATP to its precise location.
Since ATP is synthesized by the mitochondria and diffuses, ATP would be concentrated around the mitochondria and its concentration would decrease as the distance from the mitochondria increases \cite{matsumoto_intermitochondrial_2022}.
Mitochondrial positioning is then essential for the proper transport of ATP to its precise location in the cell, but the physical mechanism for this is still unknown.

The importance of mitochondrial positioning may become more critical as the size of the cell increases.
In particular, the nerve axons of neurons in animals are quite long, reaching lengths of centimeters in rodents and meters in large mammals \cite{rishal_axonsoma_2014}, while the size of the cell bodies of neurons is typically between a few and several tens of micrometers in diameter \cite{beebe_extracellular_2016}.
Thus, the positioning of mitochondria within a nerve axon may be essential for the distribution of ATP throughout the axon.
Indeed, it has been reported that mitochondria are aligned at nearly equal intervals within a micrometer-to-centimeter-length nerve axon \cite{miller_axonal_2004, matsumoto_intermitochondrial_2022}.
Mitochondrial alignment requires that mitochondria move away from each other.
As for the movement itself, cell biological observations have shown that mitochondria in nerve axons move by axonal transport of kinesin and dynein on microtubules \cite{nangaku_kif1b_1994, pilling_kinesin-1_2006, hirokawa_molecular_2010, sheng_mitochondrial_2012}.
However, little is known about how the repulsive movement occurs.

In this letter, we show that, contrary to intuition, direct repulsion between mitochondria is not necessary, and that mitochondrial alignment can arise only from mitochondrial ATP production and ATP concentration-dependent fluctuations in movement.
This may seem strange at first, but as mitochondria approach each other, the increase in local ATP concentration leads to an increase in motion fluctuations and effective repulsion between mitochondria, and then the mitochondria are aligned in a steady state.
This mechanism of generating an effective unidirectional force is very similar to the mechanism of the Soret effect \cite{ludwig_difusion_1856, soret_sur_1879, maeda_thermal_2011}, diffusion phoresis \cite{anderson_transport_1986, anderson_colloid_1989, ramm_diffusiophoretic_2021}, or chemophoresis \cite{sugawara_chemophoresis_2011, vecchiarelli_propagating_2014, sugawara_chemophoresis_2022}, where a force is generated that moves particles according to a gradient of temperature, diffusion constant, or adsorptive substance, respectively.
This effective unidirectional force, generated only by non-directional fluctuations in motion, is called the thermodynamic force.
Our study shows that even when mitochondria are driven by a non-directional non-equilibrium force, an effective unidirectional force can be generated by a mechanism similar to the thermodynamic force, and then mitochondrial alignment is achieved.


\begin{figure}[tbhp]
\centering
\includegraphics[width=.95\linewidth]{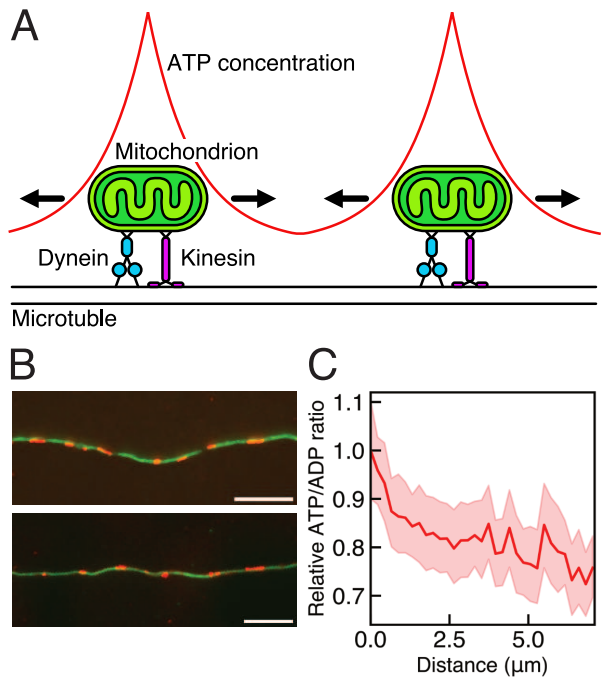}
\caption{
(A) Schematic representation of the model.
We consider one-dimensional dynamics of mitochondria and ATP concentration.
Mitochondria produce ATP, and then the gradient of ATP concentration is formed around the mitochondria.
Molecular motors, i.e. dynein and kinesin, attach to mitochondria and move on microtubules depending on the ATP concentration.
We do not explicitly include the molecular motors in the model and instead represent the mitochondrial movements as a function of ATP concentration.
(B) Microscopic images of axons showing mitochondria and tubulin stained with MitoTracker Red CM-H2XRos (red) and anti-tubulin antibody (green), respectively.
Scale bars represent 10 $\mu$m.
(C) Relative ATP:ADP signal ratio at different distances from a mitochondrion along the axon.
The red line and red band represent the mean and 95\% confidence interval, respectively.
This figure was adopted from \cite{matsumoto_intermitochondrial_2022}.
}
\label{fig:model}
\end{figure}

Here we consider the movement of mitochondria along one-dimensional microtubules in a nerve axon (Fig. \ref{fig:model}).
Many molecular motors, i.e., dynein and kinesin, move along the microtubules by using chemical energy from ATP \cite{vale_way_2000, hirokawa_molecular_2010}.
Although dynein and kinesin move in opposite directions \cite{hancock_bidirectional_2014} and have different properties, we assume that they have the same property for simplicity.
Since mitochondria are sufficiently large, the thermal noise of their motion is negligible. 
Instead, mitochondria attach and detach stochastically to molecular motors moving forward or backward, and if we observe their motion on a slower timescale than that of attachment and detachment, mitochondria appear to exhibit a random walk.
Furthermore, when a mitochondrion detaches from a molecular motor, it does not move, and we cannot observe the inertia of the movement.
We then model the mitochondrial motion as a random walk using the overdamped Langevin equation \cite{gardiner_stochastic_2009}.

Molecular motors can only move if they are attached to an ATP molecule \cite{vale_way_2000, hirokawa_molecular_2010}, and then the probability of movement increases with the concentration of ATP.
Just as ambient temperature determines the intensity of Brownian motion, ATP concentration determines the intensity of mitochondrial motion.
We consider the case where there is no explicit force to align the mitochondria.
Therefore, the Langevin equation for the position of the $i$th mitochondrion is given by
\begin{align}
\frac{dx_i}{dt} &= f \left( a(x_i) \right) \eta_i (t), \label{eq:mitochondrial_motion} \\
\langle \eta_i (t) \rangle &= 0, \nonumber \\
\langle \eta_i (t) \eta_j (t') \rangle &= 2 \delta_{i,j} \delta (t - t'), \nonumber
\end{align}
where $a(x_i)$ is the ATP concentration at $x_i$, $f(a)$ is an increasing function of $a$ because the probability of moving increases with the concentration of ATP.

Here we consider the concentration of ATP around the mitochondria.
It is natural to assume that the diffusion and consumption of ATP is much faster than the movement of the mitochondria, because ATP is a small molecule and is consumed by too many molecules.
Then we assume that the ATP concentration immediately relaxes to the steady state value following the mitochondrial movement.
ATP is produced by mitochondria, consumed as an energy source, and diffuses.
We assumed that the concentrations of molecules consuming ATP are uniformly distributed in space, and then the consumption of ATP is spatially uniform.
Thus, if the consumption of ATP is linearly proportional to its concentration due to mass action, the equation for the ATP concentration $a$ produced by a mitochondrion located at $x = x_i$ is
\begin{equation}
\frac{\partial a(x, t)}{\partial t} = p_a \delta (x - x_i) + D_a \frac{\partial^2 a}{\partial x^2} - d_a a, \label{eq:ATP_diffusion}
\end{equation}
where $p_a$, $D_a$, $d_a$ are the production rate, diffusion constant, and consumption rate of ATP, respectively.
When the ATP concentration at $x = \infty$ and $-\infty$ is 0, the steady state ATP concentration produced by one mitochondrion is
\begin{equation}
a_i^* = a_0 e^{-\sqrt{\frac{d_a}{D_a}} \left|x - x_i\right|},
\end{equation}
where $a_0$ is given by $a_0 = p_a / d_a$.
This is consistent with the previous experimental observation (see Fig. 1C and \cite{matsumoto_intermitochondrial_2022}).
Since Eq. \ref{eq:ATP_diffusion} is linear, the concentration of ATP produced by different mitochondria can be linearly superposed, and the ATP concentration is given by
\begin{equation}
a^* = a_0  \sum_{i=1}^{N} e^{-\sqrt{\frac{d_a}{D_a}} \left|x - x_i\right|},
\end{equation}
where $N$ is the number of mitochondria.


First, we show that there is an effective repulsive force between mitochondria.
For simplicity, we consider the interaction between two mitochondria, one of which is fixed at $x = 0$.
If the position of the freely moving mitochondrion is $x = x_1$, the ATP concentration at this location is given as the sum of the ATP produced by the fixed and moving mitochondria as $a(x_1) = a_0  + a_0 \exp \left( -\sqrt{\frac{d_a}{D_a}} \left| x_1 \right| \right)$.
Then, the dynamics of the freely moving mitochondrion is given by
\begin{equation}
\frac{dx_1}{dt} = f \left( a_0 \left(1 + e^{-\sqrt{\frac{d_a}{D_a}} \left| x_1 \right|} \right) \right) \eta_1 (t).
\end{equation}
From the above equation, we solve the Fokker-Planck equation by considering the above equation a the Stratonovich stochastic differential equation as
\begin{widetext}
\begin{align}
\frac{dP(x_1, t)}{dt} &= \frac{\partial}{\partial x_1} \left[ f \left( a(x_1) \right) \frac{\partial}{\partial x_1} \left\{ f\left( a(x_1) \right) P(x_1, t) \right\} \right] \nonumber \\
&= -\frac{d f(a)}{da} \frac{\partial}{\partial x_1} \left\{ \frac{d a(x_1)}{d x_1} f\left( a(x_1) \right) P(x_1, t) \right\} + \frac{\partial^2}{\partial x_1^2} \left\{ f\left( a(x_1) \right) P(x_1, t) \right\} \nonumber \\
&= \pm a_0 \sqrt{\frac{d_a}{D_a}} \frac{d f(a)}{da} \frac{\partial}{\partial x_1} \left\{ e^{-\sqrt{\frac{d_a}{D_a}} \left| x_1 \right|} f\left( a(x_1) \right) P(x_1, t) \right\} + \frac{\partial^2}{\partial x_1^2} \left\{ f\left( a(x_1) \right) P(x_1, t) \right\}, \label{eq:repulsive_force_two_bodies}
\end{align}
\end{widetext}
where if $x_1 > 0$ or $< 0$, a sign of the first term is positive or negative, respectively.
The first and second terms in the above equation indicate an anisotropic flow and an isotropic diffusion, respectively.
Although no explicit force is added to the mitochondrion as in Eq. (\ref{eq:mitochondrial_motion}), there is an effective force between mitochondria due to the existence of an ATP concentration gradient similar to the thermodynamic force.
Here, $f(a)$ is an increasing function of $a$, and then $\frac{d f(a)}{da}$ is positive.
Hence, the first term works by increasing the distance between two mitochondria, i.e., the active thermodynamic force between two mitochondria works as a repulsive force.

\begin{figure}[tbhp]
\centering
\includegraphics[width=.95\linewidth]{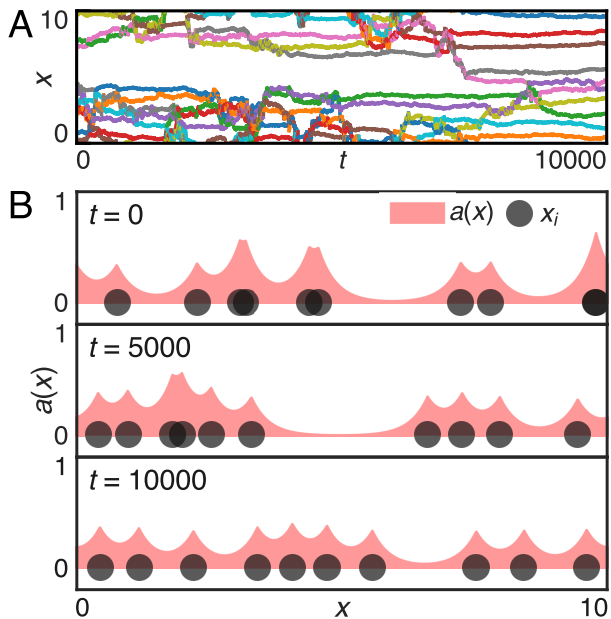}
\caption{
Time evolution of mitochondrial positions.
(A) Each line corresponds to the time evolution of the position of a different mitochondrion.
At $t = 0$, the position of each mitochondrion was randomly distributed.
(B) Snapshots of mitochondrial positions and ATP concentration.
The black circles are the mitochondrial positions, and the red area is the ATP concentration.
We set $N$ to 10, $L$ to 10, $a_0$ to 0.3, and $d_a/D_a$ to 9.0.
We numerically solved the dynamics under the periodic boundary condition. See also the Supplementary Movie \cite{supp}.
\label{fig:time_dependence}
}
\end{figure}

We confirmed that this repulsive force can drive the alignment of mitochondria.
As shown in Fig. \ref{fig:time_dependence}, initially all mitochondria are randomly distributed.
When the positions of two or more mitochondria were close, the local ATP concentration around each mitochondrion was increased, and these positions fluctuated intensely.
Then, these mitochondria moved away from each other with high probability, and after they separated and isolated, the ATP concentration around a mitochondrion decreased (see also the Supplementary Movie \cite{supp}).
Therefore, the fluctuation of the position of the mitochondria also decreased, and the mitochondria were kept apart for a long time.
That is, the mitochondria moved away from each other autonomously and were almost equally spaced without the explicit repulsion force between them.

\begin{figure}[tbhp]
\centering
\includegraphics[width=.95\linewidth]{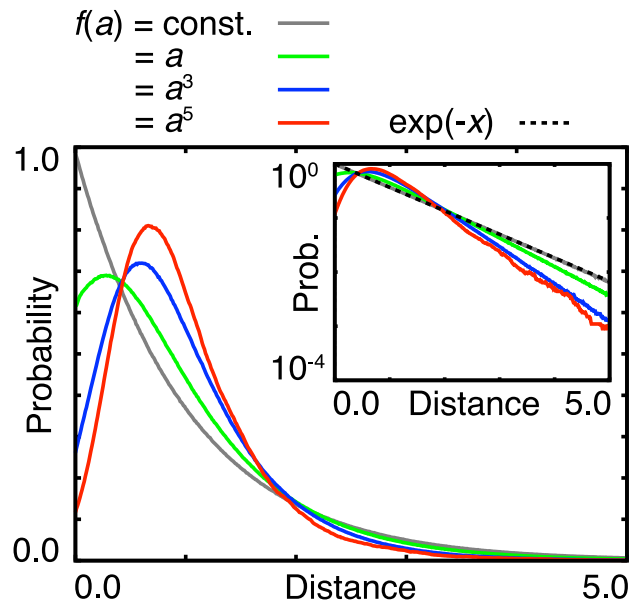}
\caption{
Probability distribution of the distance between two adjacent mitochondria at steady state.
The gray, magenta, cyan and green solid lines are for cases where $f(a)$ is given as constant, $a$, $a^3$, $a^5$, respectively.
The black dashed line is the analytically derived distribution without interaction between mitochondria.
To obtain the distribution, we set $N$ to 100 and $L$ to 100, the same ratio $N/L$ as in Fig. \ref{fig:time_dependence}.
Other parameters are the same as in Fig. \ref{fig:time_dependence}.
Inset: The distribution is plotted with a logarithmic scale.
\label{fig:prob}
}
\end{figure}

Statistical properties also showed that mitochondria move away from each other without explicit repulsive forces.
When there is no interaction between mitochondria, their position is completely random.
Then the distribution of the distance between two neighboring mitochondria in steady state follows the exponential distribution:
In the limit of infinite system size, keeping the ratio $N/L$, where $L$ is the system size, the frequency of mitochondrial existence is described by a Poisson process, where on average $N x / L$ mitochondria appear for every fixed distance $x$.
Thus, a probability distribution of the distance between two adjacent mitochondria at steady state is given by $N \exp(-N x/ L) / L$, where the probability decreases monotonically with $x$.
In fact, if $f(a)$ is given as a constant independent of $a$, i.e., each mitochondrion exhibits the random walk independent of the ATP concentration, the distribution of the distance between two adjacent mitochondria was well fitted to the exponential distribution (see gray and black dashed lines in Fig. \ref{fig:prob}), and no peaks appear at points where the distance is not zero.

In contrast, when $f(a)$ was given as an increasing function of $a$, i.e. the intensity of the fluctuation of the mitochondrial position depended on the local ATP concentration, the distribution of the distance between two adjacent mitochondria at steady state was no longer fitted by the exponential function, but showed the peak at points where the distance is not zero.
Although the peak still appeared when $f(a)$ was a linear function of $a$, the peak was less obvious because its position was close to the origin and its height was not as high (see magenta line in Fig. \ref{fig:prob}).
The peak appeared more obvious when $f(a)$ was a higher order equation for $a$ with the same parameter set (see cyan and green lines in Fig. \ref{fig:prob}).
This is because the higher the order of $f(a)$ for $a$, the greater the difference in the magnitude of the fluctuations compared to when a mitochondrion exists alone or in the vicinity of other mitochondria.
Thus, the thermodynamic repulsive force between mitochondria will also be greater since $f(a)$ is of higher order.
Indeed, the repulsive force between two mitochondria is greater if $f(a)$ is of higher order than $a$, as can be seen from Eq. (\ref{eq:repulsive_force_two_bodies}).
In any case, if the magnitude of the fluctuations in mitochondrial position depends positively on the ATP concentration, the mitochondria will align at steady state even if there is no explicit repulsive force between them.

Here we show that non-directional fluctuations in mitochondrial movement and mitochondrial ATP production are sufficient to align mitochondria along a nerve axon.
This suggests that mitochondrial function itself is linked to patterning, and that no special function such as signaling between mitochondria is required for equispaced patterning.
In addition, it has been observed experimentally that the movement of mitochondrial position in nerve axons is initially strong, but as it gradually approaches the iso-pattern, the movement becomes weaker \cite{matsumoto_intermitochondrial_2022}, as we observed in Fig. \ref{fig:time_dependence}.
For this gradual fixation, the nonlinear dependence of the fluctuation of the mitochondrial movement on the ATP concentration will play an important role;
The stronger the nonlinearity of the dependence of the fluctuations on the ATP concentration, the greater the relative difference in the fluctuations when the mitochondria are close together and when they are far apart, and the less likely they are to move when the mitochondria are aligned.
Since multiple molecular motors are known to act in concert for cargo transport \cite{miller_cross-bridges_1985, ashkin_force_1990, klumpp_cooperative_2005, hancock_bidirectional_2014}, the ATP concentration dependence of mitochondrial fluctuations would inevitably be nonlinear.
In the future, such nonlinearity will be validated both experimentally and theoretically using more microscopic models.
Note that even after alignment, the fluctuations continue, albeit weakly.
Therefore, in the real system, after the initial alignment by a mechanism we propose here, there may be a mechanism to further fix the position of the mitochondria.
Indeed, some mechanisms have been proposed to arrest mitochondria after positioning \cite{kang_docking_2008, sheng_mitochondrial_2012, schwarz_mitochondrial_2013}.

Mitochondrial alignment is thought to be physiologically important for maintaining a uniform ATP concentration in cells \cite{matsumoto_intermitochondrial_2022}.
The mechanism proposed here shows that mitochondria move to autonomously resolve deviations in ATP concentration without any special mechanism.
Moreover, it has long been known that mitochondria move in the direction of lower ATP concentrations in a nerve axon \cite{morris_regulation_1993}.
Our results are in good agreement with this observation.
Although we consider a one-dimensional system here because the nerve axon is pseudo-one-dimensional, the mechanism presented will also work for higher-dimensional systems.
Therefore, the proposed mechanism can be used to achieve a uniform intracellular ATP concentration in different cells, even beyond the nerve axon.
Furthermore, a similar mechanism may work for the uniform distribution of organelles other than mitochondria and molecules.
This study will provide an important basis for future discussions of the distribution of substances within cells.

In this letter, we have shown that nondirectional motion due to non-equilibrium processes, not thermal noise, can generate unidirectional motion of biological matter.
This is more like thermodynamic forces driven by thermal fluctuations than the motion of active matter, and we term this active thermodynamic force.
In fact, Eq. (\ref{eq:mitochondrial_motion}) contains neither interaction terms between mitochondria nor ATP gradient-dependent terms, unlike many equations describing active matter \cite{vicsek_novel_1995, marchetti_hydrodynamics_2013}.
This fundamental equation contains only the ATP concentration-dependent fluctuation like the Brownian particle in the thermal gradient.
Although the equations governing motion are so simple, by combining them with non-equilibrium reactions that change the chemical field, we found that interactions between biological matter can occur through the active thermodynamic force.
This result reminds us of the diversity of mechanisms governing the motion of biological matter.
Our study will be a pioneering step in understanding the motion of biological materials due to active thermodynamic forces driven by nonequilibrium processes, and it is expected that both experiment and theory will reveal in further studies that a variety of processes are driven by similar mechanisms.

\begin{acknowledgments}
We would like to thank Kunihiko Kaneko, and Shuji Ishihara for fruitful discussion. This work was partially supported by the JSPS KAKENHI under Grant Numbers 20K06889, 21K15048, and 21K17851.
\end{acknowledgments}




%

\end{document}